\documentclass[conference]{IEEEtran}

\usepackage{
	hyperref,
	graphicx,
	amssymb,
	algorithm,
	algorithmicx,
}
\usepackage[noend]{algpseudocode}

\newcommand*\Let[2]{\State #1 $\gets$ #2}
\algrenewcommand\algorithmicrequire{\textbf{Precondition:}}
\algrenewcommand\algorithmicensure{\textbf{Postcondition:}}

\newcommand{\backtrackarrow}{\rotatebox[origin=c]{-90}{$\curvearrowleft$}}

\newcommand{\complexity}[9]{
\begin{table}[ht]
	\begin{center}
		\begin{tabular}{l|c|c}
			\textbf{Operation} & \textbf{#1} & \textbf{#2} \\
			\hline
			\textit{StartTimer} & $O(#3)$ & $O(#4)$ \\
			\textit{PerTick} & $O(#5)$ & $O(#6)$ \\
			\textit{DeleteTimer} &  $O(#7)$ & $O(#8)$ \\
		\end{tabular}
		\linebreak
		\caption{Runtime Complexity for #9}
	\end{center}
\end{table}
}

\title{Timer Lawn: Unbound Low Latency Timer Data Structure for Large Scale, High Throughput Systems}

\author{
	\textsc{Adam Lev-Libfeld} \\[1ex]
	\normalsize Tamar Labs \\
	\normalsize Tel-Aviv, Israel \\
	\normalsize{adam@tamarlabs.com}
}
\date{\today}

\begin{document}

\maketitle

\begin{abstract}
As the usage of Real-Time applications and algorithms rises, so does the importance of enabling large-scale, unbound algorithms to solve conventional problems with low to no latency becomes critical for product viability\cite{REND, WHT}. Timer algorithms are prevalent in the core mechanisms behind operating systems\cite{BSD}, network protocol implementation, real-time and stream processing, and several database capabilities. This paper presents a field-tested algorithm for low latency, unbound timer data structure, that improves upon the well excepted Timing Wheel algorithm. Using a set of queues mapped to by TTL instead of expiration time, the algorithm allows for a simpler implementation, minimal overhead no overflow and no performance degradation in comparison to the current state of the art.
\end{abstract}

\begin{IEEEkeywords}
Stream Processing, Timing Wheel, Dehydrator, Callout facilities, protocol implementations, Timers, Timer Facilities, Lawn.
\end{IEEEkeywords}

\section{Introduction}
This paper presents a theoretical analysis of a timer data-structure designed for use with hi-throughput computer systems called Lawn. In this paper, it will be shown that although the current state of the art algorithm is theoretically optimal, under some use cases (namely where max TTL is unpredictable, or the needed Tick resolution may change) it is under-performing due to the overflow problem, which the algorithm presented here addresses. Utilizing Lawn may assist in improving overall performance and flexibility in TTL and Tick resolution with no need for any prior knowledge of the using system apart from it not utilizing non-discrete stochastic values for timer TTLs.
\subsection{Model}
In a similar manner to previous work\cite{TW, CQ, EMP}, the model discussed in this paper shall consist of the following components, each corresponding with a different stage in the life cycle of a timer in the data store:

\begin{center}
	$
	StartTimer \linebreak 
	\downarrow \linebreak  
	PerTickBookkeeping \hspace{1mm} \backtrackarrow \linebreak 
	\swarrow  \hspace{15pt}  \searrow \linebreak 
	DeleteTimer \hspace{15pt} TimerExpired
	$
\end{center}

\paragraph{\textbf{StartTimer(TTL,timerId,Payload)}} This routine is called by the client to start a timer that will expire in after the TTL has passed. The client is also expected to supply a \textit{timer ID} in order to distinguish it from other timers in the data store. Some implementations also allow the client to provide a \textit{Payload}, usually some form of a callback action to be performed or data to be returned on timer expiration.

\paragraph{\textbf{PerTickBookkeeping()}} This routine encompasses all the actions, operation and callbacks to be performed as part of timer management and expiration check every interval as determined by the data store granularity. Upon discovery of an outstanding timer to expire \textit{TimerExpired} will be initiated by this routine.

\paragraph{\textbf{DeleteTimer(timerId)}} The client may call this utility routine in order to remove from the data store an outstanding timer (corresponding with a given \textit{timer ID}), this is done by calling \textit{TimerExpired} for the requested timer before \textit{PerTickBookkeeping} had marked it to be expired.

\paragraph{\textbf{TimerExpired(timerId})} Internally invoked by either \textit{PerTickBookkeeping} or \textit{DeleteTimer} this routine entails all actions and operations needed in order to remove all traces of the timer corresponding with a given \textit{timer ID} from the data store and invoking the any callbacks that were provided as \textit{Payload} during the \textit{StartTimer} routine.

\subsection{callback run-time complexity}

Since payload and callback run-time complexity varies significantly between different data store implementations, the store of such data can be achieved for $ O(1) $ using a simple hash map, and the handling of such callbacks can be done in a discrete, highly (or even embarrassingly) parallel, this paper will disregard this aspect of timer stores.

\section{Current Solutions}

\subsection{List and Tree Based Schemes}

\complexity{List Based}{Tree Based}{n}{log(n)}{1}{1}{n}{log(n)}{common data structure based schemes}

Being included as an integral part of almost any modern programming language, these basic data structures enable convenient and simple addition of timer management to any software. That said, such simple structures suffer from oversimplification and are appropriate for very unique use cases - where the number of timers it fairly small or the ticks are far enough from one another. Using such implementations for large scale applications will require the grouping of timer producers and consumers into groups small

\subsection{Hashed Timing Wheel}
\complexity{Worst}{Mean}{n}{1}{n}{1}{1}{1}{the Hashed Timing Wheel scheme}
The Hashed Timing Wheel was designed to be an all-purpose timer storage solution for a unified system of known size and resolution\cite{TW87, TW, TWI, largescale, autoscale}. While previous work has shown that Hashed Timing Wheels have optimal run-time complexity, and in ideal conditions are in fact, optimal, real-world implementations would suffer from either being bound by maximal TTL and resolution combination, or would require a costly ($O(n)$) run-time re-build upon of the data structure upon reaching such limits (in \cite{CQ} it is referred to as "the overflow problem"). For large numbers of timers, producers, or consumers as is common in large scale operations, the simplest and mose effective solution is to overestimate the needed resolution and/or TTL so to abstain from rebuilding for as long as possible. 

\section{The Lawn Data Structure}
 
\subsection{Intended Use Cases}

This algorithm was first developed during the writing of a large scale, Stream Processing geographic intersection product\cite{VUSR} using a FastData\cite{GP} model. The data structure was to receive inputs from one or more systems that make use of a very limited range of TTLs in proportion to the number of concurrent timers they use. 

\paragraph{Assumptions and Constraints}
As this algorithm was originally designed to operate as the core of a dehydration utility for a single FastData application, where TTLs are usually discrete and variance is low it is intended for use under the assumptions that: 
\begin{center}
	$ Unique\ TTL\ Count \ll Concurrent\ Timer\ Count $
\end{center}
Assuming that most timers will have a TTL from within a small set of options will enable the application of the core concept behind the algorithm - TTL bucketing. 

\subsubsection{The Data Structure}
\begin{figure}
	\centering
	\includegraphics[width=0.9\linewidth]{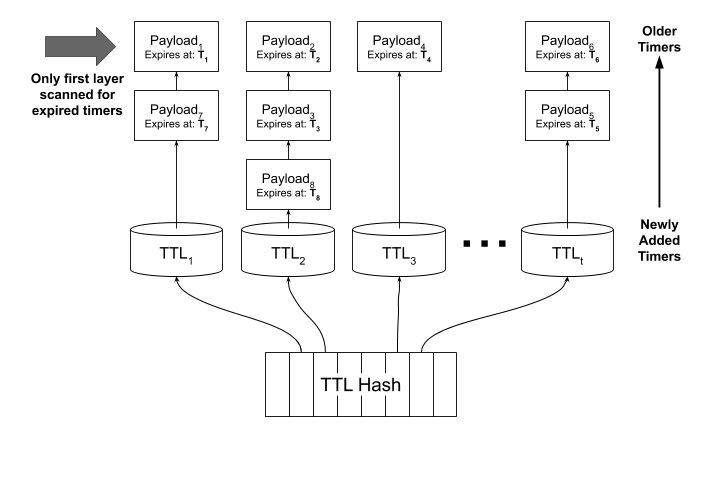}
	\caption[Lawn schematic diagram]{A schematic view of the data structure components.}
%	\Description[Lawn schematic diagram]{A schematic view of the data structure components.}
	\label{fig:Lawn1}
\end{figure}

Lawn is, at its core, a hash of sorted sets\footnote{These are the TTL 'buckets'}, much like Timing Wheel. The main difference is the key used for hashing these sets is the timer TTL. Meaning different timers will be stored in the same set based only on their TTL regardless of arrival time. Within each set, the timers are naturally sorted by time of arrival - effectively using the set as a queue  (as can be seen in fig. \ref{fig:Lawn1}). Using this queuing methodology based on TTL, we ensure that whenever a new timer is added to a queue, every other timer that is already there would be expired before the current one, since it is already in the queue and have the same TTL.

The data structure is analogous to blades of grass (hence the name) - each blade grows from the roots up, and periodically (in our case every $Tick$) the overgrown tops of the grass blades (the expired timers) are maintained by mowing the lawn to the desired level (current time).

\subsection{Algorithm}
\begin{algorithm}
	\caption{The Lawn Data Store \label{alg:lawn}}
	\begin{algorithmic}[1]
		\Require{\\$id$ - a unique identifier of a timer.\\ $ttl$ - a whole product of $Tick Resolution$ representing the amount of time to wait before triggering the given timer $payload$ action.\\$payload$ - the action to perform upon timer expiration.\\\textit{current time} - the local time of the system as a whole product of $Tick Resolution$}
		
		\Statex
		\Function{InitLawn}{{}}
			\Let{TTLHash}{new empty hash set}
			\Let{TimerHash}{new empty hash set}
			\Let{$closest\ expiration$}{0}
		\EndFunction
		\Statex
		\Function{StartTimer}{$id$, $ttl$, $payload$}
			\Let{$endtime$}{current time + $ttl$} %\Comment{current time: the local time of the system as a whole product of $Tick Resolution$}
			\Let{$T$}{$(endtime,ttl,id,payload)$}
			\Let{TimerHash[$id$]}{$T$}
			\If{$ttl \notin$ TTLHash}
				\Let{TTLHash[$ttl$]}{new empty queue}
			\EndIf
			\State{TTLHash[$ttl$].\textbf{insert}($T$)}
			\If{$endtime < closest\ expiration$}
				\Let{$closest\ expiration$}{$endtime$}
			\EndIf
		\EndFunction
		\Statex
		\Function{PerTickBookkeeping}{{}}
		\If{current time $ < closest\ expiration$}
		\State{\textbf{return}}
		\EndIf
		\For{$queue \in $TTLHash}
			\Let{$T$}{\textbf{peek}($queue$)}
			\While{$T_{endtime} < $ current time}
				\State{\textbf{TimerExpired}($T_{id}$)}
				\Let{$T$}{\textbf{peek}($queue$)}
			\EndWhile
			\If{$closest\ expiration$ = 0 \newline \textbf{or} $T_{endtime} < closest\ expiration$}
				\Let{$closest\ expiration$}{$T_{endtime}$}
			\EndIf
		\EndFor
		\EndFunction
		\Statex
		\Function{TimerExpired}{$id$}
			\Let{$T$}{TimerHash[$id$]}
			\State{\textbf{DeleteTimer}($T$)} 
			\State{\textbf{do} $T_{payload}$}
		\EndFunction
		\Statex
		\Function{DeleteTimer}{$id$}
			\Let{$T$}{TimerHash[$id$]}
			\If{$T_{endtime} = closest\ expiration$}
			\Let{$closest\ expiration$}{0}
			\EndIf
			\State{TTLHash[$T_{ttl}$].\textbf{remove}($T$)} 
			\State{TimerHash.\textbf{remove}($T$)} 
			\If{TTLHash[$T_{ttl}$] is empty}
				\State{TTLHash.\textbf{remove}[$ttl$]}
			\EndIf
		\EndFunction
		
	\end{algorithmic}
\end{algorithm}

\subsubsection{Correctness \& Completeness}
%Given Timer $T$ with TTL $TTL$ that is started (using the \textit{StartTimer} operation) at time $t$,  
To prove the algorithm's correctness, it should be demonstrated that for each Timer $t$ with TTL $ttl$,  \textit{TimerExpired} operation is called on $t$ within \textit{Tick} of $ttl$. Since the algorithm pivots around the TTL bucketing concept, wherein each timer is stored exactly once in its corresponding bucket, and these buckets are independent of each other, it is sufficient to demonstrating correctness for all timers of a bucket, That is:
\begin{center}
	$\forall \quad T^{start}, T^{ttl} \in \mathbb{N} \quad \exists \quad T^{stop} \in \mathbb{N} : T^{stop}-T^{start} \approx T^{ttl}$
\end{center}

Alternatively, we can use the sorting analogy made by G. Varghese et al.\cite{TW} to show that given two triggers $T_n, T_m$: 

\begin{center}
	$\forall \quad T_n, T_m \quad | \quad T^{start}_n<T^{start}_m,  T^{ttl}_n = T^{ttl}_m \quad \exists\quad T^{stop}_n, T^{stop}_m  \Rightarrow T^{stop}_n < T^{stop}_m$
\end{center}

Taking into account that each bucket only contains triggers with the same TTL we can simplify the above:

\begin{center}
	$\forall \quad T_n, T_m \quad | \quad T^{start}_n<T^{start}_m \Rightarrow T^{stop}_n < T^{stop}_m$
\end{center}

Which, due to the bucket being a sorted set, ordered by $T^{start}$ and triggers being expired by bucket order from old to new is self-evident, and we arrived at a proof.

\subsubsection{Space and Runtime Complexity}
The Lawn data structure is dense by design, as every timer is stored exactly once, a new trigger will add at most a single TTL bucket and empty TTL buckets are always removed, the data structure footprint will only grow linearly with the number of timers. Hence, overall space complexity is linear to the number of timers ($O(n)$). 

\complexity{Worst}{Mean}{1}{1}{n}{t\sim1}{1}{1}{the Lawn scheme}

Since the \textit{PerTickBookkeeping} routine of Lawn iterates over the top item of all known TTL buckets on every expiration cycle (where at least one timer is expected to expire), it's mean case runtime is linear to $t$ (the number of different TTLs) and seems to be lacking even in comparison to more primitive implementations of timer storage. That said, with an added assumption that the TTL set size is roughly constant over time, or at worst asymptotically smaller then the number of timers, we can regard this operation as constant time.

This assumption is valid in our case as it is derived from the needs of the algorithm users, these being other computer systems, which often have either a single TTL used repeatedly, their TTLs are chosen from a list of hard-coded values or derived from a simple mathematical operation (sliding windows are a good example of this method, using fixed increments or powers of 2 to determine TTLs etc.). Computer systems which are using highly variable TTL values are suitable for usage with this timer algorithm only under specific circumstances (such as multi-worker expiration system as described below).

\paragraph{Space complexity} is $O(n)$ since at worst case each timer is stored in its own bucket alongside a single entry in the timer hash. To compare, this spatial footprint is bound from above by that of the Hashed Hierarchical Timing Wheel, as due to it's multi-level structure a single timer can be pointed at by a chain of hierarchical wheels, increasing its overall space requirement.

\section{Comparison and Reflection}
While general use systems, aggregating timers from several sources with, or applications with highly predictable needs may benefit from the relative stability of run-time provided by Timing Wheel (let alone the fact that it has been shown to be an optimal solution in terms of run-time complexity) Large scale machine serving systems would suffer from the overflow problem when faced with unpredictable scale of usage. This is handled in Lawn by a "slow and steady" approach, optimizing for specific use cases.

Designed for large scale, high throughput systems, Lawn has displayed beyond state of the art performance for systems complying with its core assumptions of a multi-worker, hi-frequency, hi-timer-count with low TTL variance applications.

\begin{table}[ht]
	\begin{center}
		\begin{tabular}{l|c|c}
			%& \multicolumn{2}{c}{\textbf{Case}} \\
			\textbf{Operation} & \textbf{Timing Wheel} & \textbf{Lawn} \\
			\hline
			\textit{StartTimer} & $O(1)$ & $O(1)$ \\
			\textit{PerTick} & $O(1)$ & \boldmath{$O(t\sim1)$} \\
			\textit{DeleteTimer} & $O(1)$ & $O(1)$ \\
			\textit{TimerExpired} & $O(1)$ & $O(1)$ \\
			\textit{overflow} & \boldmath{$O(n)$} & $O(1)$ \\
			\textit{space} & $O(n)$ & $O(n)$ \\
		\end{tabular}
		\linebreak
		\caption{Mean Runtime Complexity comparison}
	\end{center}
\end{table}

\subsection{A View of Multiprocessing} 
Unlike the state-of-the-art Timer Wheel algorithm, Lawn enables the simultaneous timer handling and bookkeeping by splitting the buckets between several worker processes or threads, adding or removing workers as needed. This method enables usage of the Lawn algorithm in highly parallel applications and does not require the use of semaphores than other synchronization mechanisms within the bucket level.

\subsection{Known implementations of Lawn}
As mentioned in the body of this paper, the Lawn algorithm has already been tested and deployed in several programming languages by different organizations. Some of these implementations were developed by or in tandem with the author of this paper and some with his permission all with reported improvement in performance. The algorithm is free to use and the source code for many of these implementations has been published under an open source license.
\begin{enumerate}
	\item Redis Internals \cite{Redis} - a high performance in-memory key-value store - uses Lawn implementation for streams and other internal timers.
	\item ReDe event dehydrator Redis module\cite{REDE}.
	\item Mellanox RDMA timers for an undisclosed Infiniband subsystem.
	\item User specific rate limiting-timers for client device power consumption optimization\cite{VUSR}.
	\item $clib$ Timer management utility lib.
\end{enumerate}

\section{Conclusion}
Lawn is a simplified overflow-free algorithm that displays near-optimal results for use cases involving many (millions) concurrent timers from large scale (tens of thousands) of independent machine systems. 
The algorithm is currently deployed and in use by several organizations under real-world load, all reporting satisfactory results.

\bibliographystyle{ieeetr}
\bibliography{lawn}

\end{document}